\documentclass[preprint,prb,preprintnumbers,amsmath,amssymb]{revtex4}
                                                                                
\usepackage{graphicx}
\usepackage{dcolumn}
\usepackage{bm}
\usepackage{ae}
\usepackage{aecompl}
\usepackage{amsmath}
\usepackage{amssymb}
\usepackage{epsfig}

\begin{document}
\title{ The phase diagram of water at high pressures as obtained by computer simulations of the TIP4P/2005 model:
the appearance of a plastic crystal phase. }
{\small Physical Chemistry Chemical Physics, volume 11, 543-555, (2009)}                                                                                
                                                                                
                                                                                
\author{J.L.Aragones, M.M.Conde, E.G.Noya and  C.Vega  }
\affiliation{Dpto. de Qu\'{\i}mica F\'{\i}sica, Facultad de Ciencias
      Qu\'{\i}micas,  Universidad Complutense, 28040 Madrid, Spain}
                                                                                
\date{\today}
\begin{abstract}

 In this work the high pressure region of the phase diagram of water has been studied by 
computer simulation by using the TIP4P/2005 model of water. Free energy calculations 
were performed for ices VII and VIII and for the fluid phase to determine the melting 
curve of these ices. In addition molecular dynamics simulations were performed at 
high temperatures (440K) observing the spontaneous freezing  of the
liquid into a solid phase at pressures of about 80000 bar. 
 The analysis of the structure obtained lead 
to the conclusion that a plastic crystal phase was formed. In the plastic crystal  phase
the oxygen atoms were arranged forming a body center cubic structure, as in ice VII, but 
the water molecules were able to rotate almost freely. 
Free energy calculations were performed for this new phase, and it was found that for
TIP4P/2005 this plastic crystal phase is  thermodynamically stable with respect to 
ices VII and VIII  for temperatures higher than about 400K, although the precise value depends on the pressure. 
By using Gibbs Duhem simulations, all coexistence lines were determined, and the  
phase diagram of the TIP4P/2005 model was obtained, including ices VIII and VII and 
the new plastic crystal phase.  The TIP4P/2005 model 
is able to describe qualitatively the phase 
diagram of water. 
It would be of interest to study if such a plastic crystal  phase does 
indeed exist for real water.
The nearly spherical shape
of water makes possible the formation of a plastic crystal phase at high temperatures.
The formation 
of a plastic crystal  phase at high temperatures (with a bcc arrangements of oxygen atoms) is 
fast from a kinetic point of view occurring in about 2ns. This is in contrast to the nucleation 
of ice Ih which requires simulations of the order of hundreds of ns. 
\end{abstract}

\maketitle

  \section{INTRODUCTION} 
  The study of the phase diagram of water started its way with the pioneering work 
of Tammann\cite{tammann,tammann1910} and Bridgman\cite{bridgman12,bridgman35,bridgman37}, who obtained a number of different solid phases for water. 
The story has not ended yet since in the last decade new solid phases have been 
discovered, as it is the case of ice XII\cite{N_1998_391_00268}, discovered in 1998 by 
Lobban, Finney and Kuhs  and ices XIII and 
XIV discovered just two years ago by Salzmann 
et al.\cite{iceXIII}. In addition to the existence of new phases, 
water presents a number of interesting questions, as the possibility of a liquid-liquid
transition\cite{N_1992_360_00324_nolotengo,mishima98}, the relation between the 
different amorphous phases\cite{debenedetti03}, the possibility 
of partial proton ordering\cite{JCP_2000_112_07169} and the aspect of the phase diagram at high temperatures 
and pressures. 
 Water is an important molecule since its liquid
phase constitutes the "matrix of life" \cite{ball01,franksbook}.  
Getting a molecular understanding of the
behaviour of water\cite{finneyreview} appears as an interesting and challenging goal. 
For this reason thousands of computer simulation studies have been reported since
the pioneering work of Barker and Watts\cite{barker69} and Rahman 
and Stillinger\cite{rahman71}. Most of the studies
have focused on the liquid phase. The study of the solid phases have received less 
attention. Morse and Rice\cite{morse82} analyzed the performance of early water models to describe
the different solid phases of water. In the recent years we have analysed the performance
of several popular models (SPC/E\cite{berendsen87} and TIP4P\cite{jorgensen83}) to 
describe the  solid phases
of water and its phase diagram\cite{sanz1,vegafilosofico,vegautrecht,mcbride1}. It was 
realized that TIP4P could be modified slightly
to yield an overall better description of the water phase diagram. 
In this way we proposed two new models
TIP4P/2005\cite{abascal05b} and TIP4P/Ice\cite{abascal05a}, the first 
describing correctly the maximum in density of water
and the second reproducing the melting point temperature of ice Ih. After analysing the
behaviour of these two models for other properties (diffusion coefficients, surface tension,
vapour-liquid equilibria) it has been concluded that TIP4P/2005 provides a quite reasonable
description of the properties of water. In fact it reproduces reasonably well  
the densities and relative energies of the 
water polymorphs\cite{aragones1,noya_jpc_2007,maria1}, the vapour-liquid equilibria\cite{vega06}, the diffusion 
coefficient\cite{abascal05b}, the surface tension\cite{vega07a} and the structure of the liquid and solid phase (as given 
by the radial distribution functions). In the original paper where TIP4P/2005 was proposed 
it was shown that it provides a reasonable description of the phase diagram of 
water\cite{abascal05b} at low and moderate pressures where the polymorphs Ih, II, III, V and VI are
thermodynamically stable. However, the high pressure and high temperature region was not considered.
One of the goals of this paper is to determine the high temperature and the high pressure part of 
the phase diagram including the high pressure polymorphs, namely ices VII and VIII. 

 Another issue which motivated this research was the possibility of obtaining by computer 
simulation the nucleation of ice from liquid water.  
In principle, by cooling water at room pressure in a computer simulation, there is the possibility of 
nucleating ice Ih (or Ic). Svishchev and Kusalik\cite{svishchev94} 
were able to nucleate
ice from liquid water by using an external field. Attempts to obtain ice from water without 
external fields were 
unsuccessful  until the seminal work of Matsumoto, Saito and Ohmine \cite{matsumoto02},
who in 2002 observed by the first time the
nucleation of ice from liquid water after very long runs (hundreds of ns). After these
first attempts several other authors have been able to nucleate ice from water. Vrbka and Jungwirth\cite{subsurface}  nucleated ice 
from water in a very long run using a slab of liquid in contact with its vapour. 
Quigley and Rodger \cite{quigley_rodger} nucleated ice by introducing
bias in the simulations (via the metadynamics technique\cite{metadynamics}). However, it is fair to say that nucleating ice Ih (or Ic) from liquid 
water in a computer simulation is rather difficult, and this can be achieved only either 
performing very long runs or introducing bias within the simulations.
 However, in a very recent experimental study following previous work\cite{dolan_JCP1232005}, Dolan et al.\cite{dolan_nature_physics_2007} showed 
that the formation of ice VII, from water occurs in a few nanoseconds
 when the 
temperature was around 400K. In their work, Dolan et al.\cite{dolan_nature_physics_2007} suggested
that such low nucleation 
time could be studied by computer simulation. We intend to address this 
issue in this paper.

   In the present work we shall determine by using free energy calculations the high pressure 
region of the phase diagram of the TIP4P/2005 model. On the other hand
we shall use this model to analyse if the nucleation of ice VII from the liquid
does indeed occur in a few ns. As it will be shown the answer is positive, and
besides, as a surprise a new solid phase is found for water. This new solid phase
is a plastic crystal phase. It will be shown that for TIP4P/2005 
this plastic crystal phase is 
thermodynamically stable with respect to ices VII and VIII  
in the region of high pressures and temperatures.
The plastic crystal has also been found two months ago by 
Takii, Koga and Tanaka \cite{takii_koga_tanaka} for
the TIP5P and TIP4P models of water.  
It would be of interest to determine if this plastic crystal phase does indeed 
exist for real water, 
especially taken into account that the interest in 
the high pressure phases of water is growing\cite{hemley_nature1987,hemley_JCP1282008}. 
If so, that would be, ice XV (the precise value of the roman numeral depends on the order ices are found experimentally).

\section{ Simulation details }
 In this work Monte Carlo  simulations have been performed for 
ices VIII, VII and liquid water. The TIP4P/2005 model  
will be used. This a rigid non-polarisable model,
with a  distribution of charges, similar to that of the TIP4P model. A Lennard-Jones (LJ) 
site is located on 
the oxygen atom, positive charges are located on the hydrogen atoms, and the negative charge is located
in the H-O-H bisector. The parameters of the model are given in Table I. The TIP4P/2005 model improves the description of water properties significantly with respect to the original TIP4P 
model\cite{vega_faraday} and is becoming popular in computer simulations of water\cite{nezbeda08,hugh08,franceses08,bartok_baranyai}. A critical comparison of the performance of the model with respect to other water models has been presented recently\cite{faraday}. 

Ice VIII is formed by two sub-lattices, which are interpenetrated but not 
interconnected\cite{petrenko99,eisenberg69}. 
In one of the sublattices all dipole moments of the water molecules are aligned along the 
negative direction of the vector {\bf c} of the unit cell, whereas in the other sublattice all dipole moments
are aligned along the positive direction of the {\bf c} vector of the unit cell. Thus the net dipole 
moment of ice VIII is zero.  Protons are ordered in ice VIII. Thus an initial configuration 
of ice VIII can be obtained easily from the neutron diffraction 
experimental data\cite{JCP_1984_81_03612} 
(adjusting slightly the bond length and angles from the experimental
values to those of the TIP4P/2005 model). Ice VII is formed by 
two ice Ic sublattices which are interpenetrated but not interconnected. Ice VII is a proton 
disordered phase\cite{kamb_ice_VII,whalley_ice_VII,eisenberg69}. The initial configuration of 
ice VII was obtained by generating first 
an ice Ic lattice, with no net dipole moment and satisfying the Bernal Fowler rules\cite{bernal33,pauling35}. 
The algorithm of Buch et al.\cite{buch98} was used to obtain such a proton disordered configuration. 
A second ice Ic lattice was obtained in the same way. After these two independent 
ice Ic configurations were obtained, we generated the initial configuration of ice VII 
but putting together the two ice Ic sublattices, so that they are interpenetrated but not 
interconnected. Thus, the density of ice VII is of about twice the density of ice Ic.

 The number of molecules used in the simulations 
were 600 (ice VIII), 432 (ice VII), 360 (ice VI) 
and  for water we used either 432 or 360. In all cases the LJ 
potential was truncated at $8.5$ \AA.
Standard long range corrections to the LJ part of the potential were added
(i.e., $g(r)=$1 beyond the cutoff)\cite{allen_book}.
Note that this simple prescription (i.e. assuming $g(r)=$1 beyond the cutoff)
might introduce an error
in the estimate of the long range correction,
especially in solid phases (where g(r) is not one even at large distances).
The impact that this might have on the coexistence
points is not easy to assess, as the effect is different
for different phases and at different densities. By analyzing the internal energy 
of solid phases for different systems sizes we conclude that the use 
of the simple prescription  $g(r)=$1 to estimate the long range correction 
of the LJ part of the potential may introduce an error of about 1.5\% in 
the determination of coexistence points
,which is enough for the purposes of this work.
The importance of an adequate treatment of the long range coulombic forces when
dealing with water simulations has been pointed out in recent
studies.\cite{JCP_1998_108_10220,lisal02,rick04JCP_2004_120_06085}
In this work, the Ewald summation technique\cite{frenkelbook} has been employed for the
calculation of the long range electrostatic forces.
Isotropic NpT simulations were used for the liquid phase and for ice VII (which is 
of cubic symmetry) while anisotropic Monte Carlo
simulations (Parrinello-Rahman like)
\cite{JAP_1981_52_007182,yashonath85} were used for
ices VIII and VI (both tetragonal).
 A typical Monte Carlo run involves about 30000 cycles of equilibration and 70000 cycles
to obtain averages (defining a cycle as a trial move per particle plus a trial volume 
change). 

The free energy was computed at a certain thermodynamic state for the ices VIII, VII and for liquid
water. Once obtained at a certain thermodynamic state the free energy can be computed for any other
thermodynamic conditions by using thermodynamic integration. 
Initial coexistence points  were located by imposing the condition of equal 
chemical potential, temperature and pressure between coexistence phases. 
In particular we located coexistence points between several phases 
along the 70000 bar isobar, and along the 400K isotherm. 
The free energy of the liquid is calculated by computing the free energy
change along a reversible path in which
the water molecules are transformed into Lennard-Jones spheres by switching 
off the charges.
The free energy of the reference Lennard-Jones fluid is taken from the work of
Johnson {\it et al.} \cite{MP_1993_78_0591_photocopy}.
The free energy of ices VIII and VII was obtained by using the Einstein crystal method\cite{frenkel84},
or its slightly variant recently proposed by us denoted as the Einstein molecule method\cite{vega_noya}.
In the case of ice VII, and due to the proton disorder, one must add at the end of 
the calculation the entropy contribution due to the degeneracy of the structure. 
Thus was estimated by Pauling\cite{pauling35} to be of about $S/(Nk_{B})=ln(3/2)$. 
Further details about the free energy calculations, both for the liquid phase, and for 
the solid phases of water can be found in our recent review paper about free energy 
calculations \cite{vega_review}.  

 Once an initial coexistence point has been determined the rest of the coexistence curve 
can be obtained by using Gibbs Duhem simulations\cite{kofke93,ACP_1998_105_0405__coll_nolotengo}. 
 Gibbs Duhem simulations, first proposed by Kofke,  allow to determine the
 coexistence line between two phases, provided that an initial coexistence
 point is known.  Gibbs Duhem is just a numerical integration (using simulation
 results to estimate the volume and enthalpy change between phases) of the
 Clapeyron equation.  
  For the integration of the Clapeyron equation a fourth-order
  Runge-Kutta method algorithm is employed. 
   Anisotropic scaling (Rahman Parrinello like) was used to simulate ices VIII 
  and VI (both are tetragonal) and isotropic NpT simulations are used for the 
  liquid and for the solid phases of cubic symmetry (ice VII and the plastic crystal solid). 
  Rest of the details (size of the system, cutoff) were identical to that used in 
  the NpT simulations. 

In this work we have also performed Molecular Dynamic simulations to study the 
formation of solid phases from liquid water at high temperatures. The choice 
of Molecular Dynamics (instead of Monte Carlo) allows to determine the time 
required by the liquid to freeze so that
a comparison with experiment is possible. 
The molecular dynamic simulations were performed with the program Gromacs\cite{gromacs33}.  
In the simulations the temperature is fixed with a Nose-Hoover
thermostat\cite{nose84,hoover85} with a relaxation time of 2~ps.
To keep the pressure constant, an isotropic 
Parrinello-Rahman barostat\cite{parrinello81,nose83} was used.
The relaxation time of the barostat was 2~ps.
The time step used in the simulations was 1~fs.
The typical length of the simulations was about 5~ns (five million time steps).
The geometry of the water molecule is enforced using constraints.
The LJ part of the potential  was truncated at 8.5~\AA $\;$
and standard long
range corrections were added.
The real part of the coulombic potential was truncated at 8.5~\AA. 
The Fourier part of the Ewald sums was evaluated by using the Particle Mesh
Ewald (PME) method of Essmann {\em et al.}\cite{essmann95}
The width of the mesh was 1~\AA\ and we used a fourth other
polynomial.
As it can be seen the conditions of the Molecular Dynamic simulations 
were quite similar to those used in the Monte Carlo simulations. 

 \section{Results}

\begin{figure}[!h]
\begin{center}
\caption{MD trajectories for four different initial configurations obtained with Gromacs at 440K and 80000bar. The formation a plastic
crystal phase is indicated by the jump in the density (for two of the runs
freezing occurs at quite similar times).}
\label{figure_nucleation_ice}
\includegraphics[clip,height=0.35\textheight,width=0.60\textwidth,angle=-0]{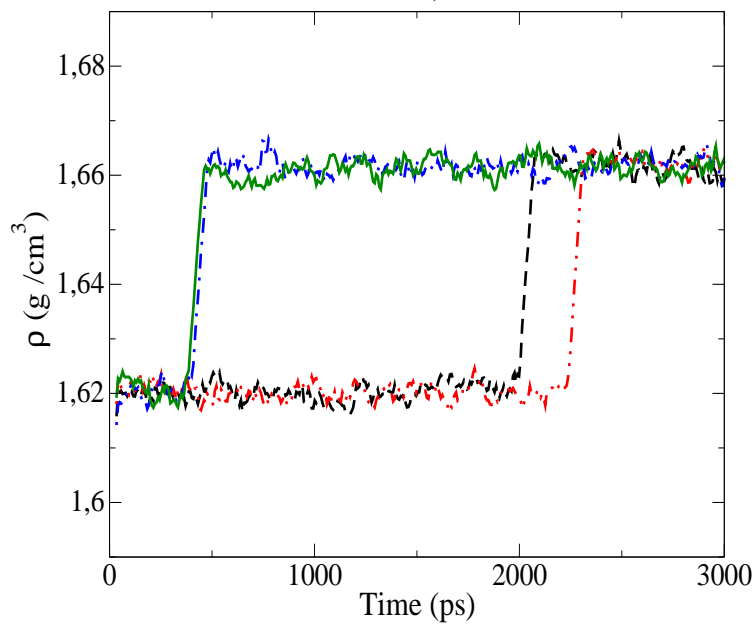}
\end{center}
\end{figure}

  Let us start by presenting the results of Molecular Dynamics for the 440K isotherm.
This temperature is close to 
temperature where fast nucleation of ice was observed experimentally\cite{dolan_nature_physics_2007}.  
 We used 432 molecules in these molecular dynamic simulations. 
The choice of this number of molecules was motivated by the fact that we were expecting 
the formation of ice VII at high pressures. This number of molecules (432) would correspond
to 216 unit cells (6x6x6) of ice VII. 
We started our simulations at low pressures (i.e about 10000 bar) where the system 
is in the liquid phase, and increased the pressure in consecutive runs.
The final configuration of a run 
was used as the initial configuration of the next run. When the pressure was of 80000 bar 
we observed the spontaneous formation of a solid phase within a few ns. We repeated the 
runs several times (using different initial configurations) and the formation of 
the solid phase in less than 5ns  was observed in almost all the cases. 
Thus the nucleation of a solid phase under  
these thermodynamic conditions can be reproduced
easily. In Fig. \ref{figure_nucleation_ice} the    
evolution of the density with time at 440K and 80000 bar is 
plotted for several independent runs (we performed five independent runs).  
The nucleation of ice is a stochastic event so 
that each run freezes in a somewhat different time. However, it is clear that 
the formation of a solid phase occurs in about 1.5ns  for the TIP4P/2005 model in 
this thermodynamic state (although for one of the five runs performed 
no freezing was observed after 5ns). 
Since the package Gromacs allows to visualise the trajectories,
it was simple to identify the structure of the solid. The oxygen atoms of the water molecules 
formed a body centred cubic structure.
The surprise was that the molecules of water were rotating  rather 
wildly in the solid phase.
Thus the system jumped from the liquid phase into a plastic crystal phase. 
Notice that
ices VIII and VII are not plastic crystal phases, so that the phase obtained spontaneously 
is a new type of solid. 
For small system sizes (N=128 and N=432) and NpT simulations
the crystal formed
tends to be commensurate with the simulation box. However, for larger system
sizes or NVT simulations we found more often that the crystal grows in some
direction that was not compatible with the simulation box, so that 
complete crystallisation could not be achieved.
In summary, we have observed that
the TIP4P/2005 model freezes spontaneously in about 1.5ns (in average) 
into a plastic crystal phase at 440K.  The formation of a plastic crystal, by heating 
ice VII at constant pressure  has also
been reported two months ago by Takii, Koga and Tanaka\cite{takii_koga_tanaka} 
for the TIP5P and TIP4P models. The plastic
crystal phase obtained in this work from the liquid is identical to that 
described by Takii, Koga and Tanaka \cite{takii_koga_tanaka}.
Sometimes when cooling/compressing liquids, the obtained solid phase is not the 
most stable from a thermodynamic point of view, but the more favourable from a kinetic 
point of view. Thus the thermodynamic stability of the plastic crystal phase with respect
to ices VIII and VII should be determined. 

\begin{figure}[!h]
\begin{center}
\caption{Oxygen-oxygen, oxygen-hydrogen and hydrogen-hydrogen
radial distribution functions at T=440K and p=80000 bar
for the bcc plastic crystal as obtained by isothermal compression of the liquid state (solid line)
and by isobaric heating of ice VII (filled circles).}
\label{fdr_plastic}
\includegraphics[clip,height=0.35\textheight,width=0.60\textwidth,angle=-0]{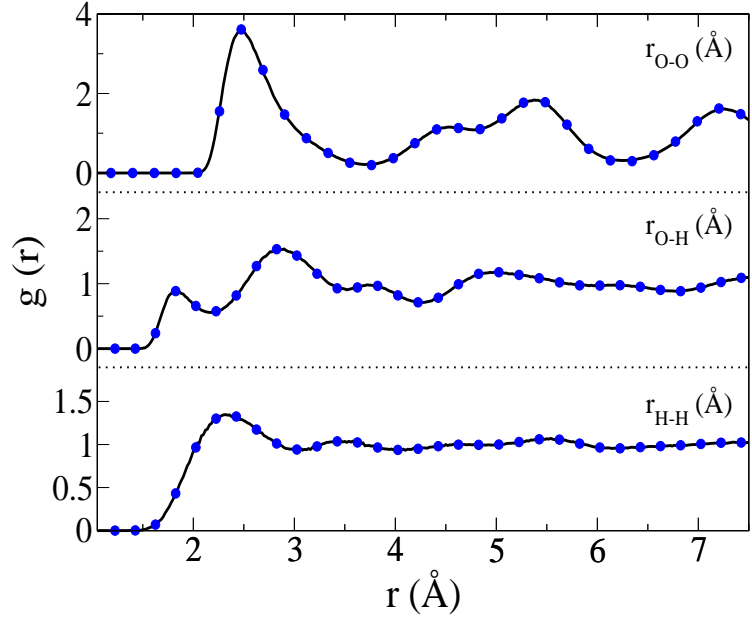}
\end{center}
\end{figure}

\begin{figure}[!h]
\begin{center}
\caption{Hysteresis loop obtained by heating ice VII to obtain the plastic crystal phase (dotted line)
and by cooling the plastic crystal phase to recover ice VII (dashed line). Results were obtained along
the 70000bar isobar. The residual part of the internal energy is plotted as a function of 
the temperature.}
\label{figure_hysteresis}
\includegraphics[clip,height=0.35\textheight,width=0.60\textwidth,angle=-0]{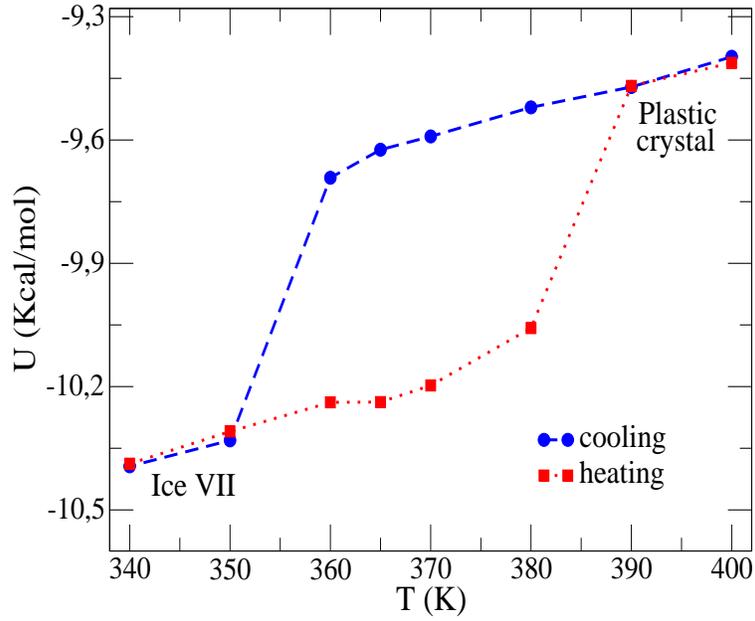}
\end{center}
\end{figure}

\begin{figure}[!h]
\begin{center}
\caption{Oxygen-oxygen, oxygen-hydrogen and hydrogen-hydrogen
radial distribution functions for ice VII (solid line), 
for ice VII obtained from the
cooling of the plastic crystal phase (filled circles) and for ice VIII 
(dashed-dotted line). The radial distribution functions
were obtained at 300K and 70000bar.}
\label{fdr_vii_viii}
\includegraphics[clip,height=0.35\textheight,width=0.60\textwidth,angle=-0]{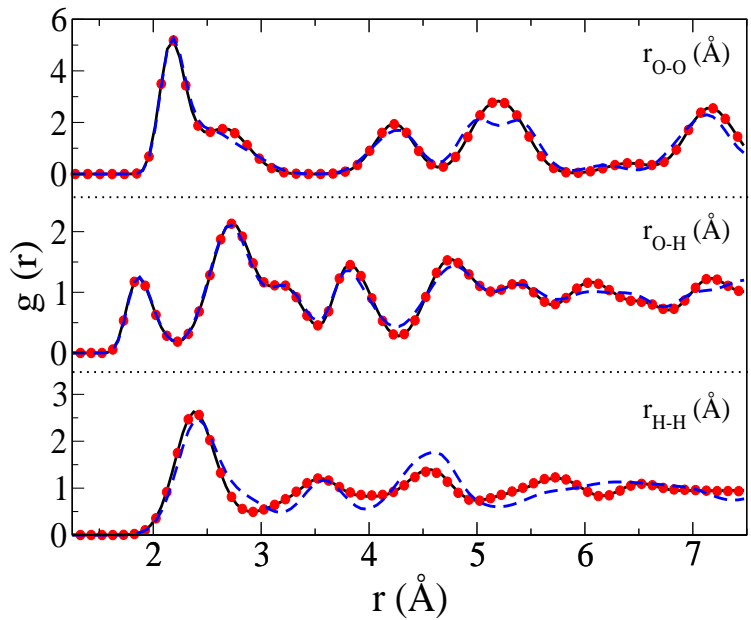}
\end{center}
\end{figure}

  After observing the formation of the plastic crystal phase we analysed in detail the 
mechanical stability of ices VII and VIII when heated along the 70000 bar isobar. Starting 
at low temperatures, the temperature was increased (in jumps of 10-20K). For each temperature
the length of the run was of about 100000 cycles. The final configuration of each simulation 
was taken as the initial configuration of the next run. For ice VII we observed a jump in the 
internal energy and density  starting at a temperature of about 380K and ending at a 
temperature of about 390K. Visualisation of several snapshots revealed that the phase obtained 
at 390K was a plastic crystal phase. The 
density, radial distribution functions and internal energy were identical to those 
of the plastic crystal phase obtained from the freezing of the liquid (when compared at the
same thermodynamic conditions).  This is illustrated for the radial distribution
functions in Fig. \ref{fdr_plastic}.
After a long equilibration at 400K and 70000 bar of the plastic crystal phase 
obtained by heating ice VII, we proceed to cool the system along the isobar 
to see if the system was able to return to the original state. The plastic crystal phase was
mechanically stable up to 360K, and at a temperature of about 350K, the density and internal 
energy undergo a jump. The result of this heating-cooling cycle is plotted in 
Fig. \ref{figure_hysteresis}. The presence of a hysteresis loop 
indicates clearly the existence of a first order 
phase transition between ice VII and the plastic crystal phase. In fact the phase obtained by 
cooling the plastic crystal phase is ice VII. In 
Fig. \ref{fdr_vii_viii}
the radial distribution 
functions (O-O, H-H and O-H) of ice VII and of the phase obtained by cooling the plastic crystal
phase are shown. As it can be seen they are nearly identical (and the same is true for 
other thermodynamic properties as density and internal energy) providing further evidence of the
fact that the solid phase obtained by cooling the plastic crystal is ice VII. By analyzing the 
average positions of the water molecules in the phase obtained from the cooling of
the plastic crystal phase, we found that the
solid  was formed by two sublattices which are interpenetrated but not interconnected as in ice VII. 
In each
of the sublattices each molecule was forming four hydrogen bonds (in two acting as donor and in
two acting as acceptor). Therefore, the phase obtained by cooling the plastic crystal phase was 
indeed ice VII.  
   We have repeated this heating-cooling cycle for the ice VII-plastic crystal transition 
 at several pressures (70000 bar, 65000 bar and 60000 bar). 
In Table II the temperatures at which the ice VII to plastic crystal and 
plastic crystal to
ice VII transitions occur are presented. 
 From the results of Table \ref{hysteresis_lopp_table} it is obvious that the ice VII to 
plastic crystal thermodynamic phase transition (where the chemical potential of both phases 
become identical) must be between 350K and 390K when the pressure is of 70000 bar, between 
340K and 380K when the pressure is 65000 bar and between 310K and 360K when the pressure is 
of 60000 bar. Obviously only free energy calculations can determine the precise location 
of the coexistence point, which must be located between the two temperatures which 
bracket the hysteresis loop. 
 
We also analysed the mechanical stability of ice VIII when heated along these three isobar. 
It was found that ice VIII was mechanically stable up to 390K for the p=70000bar isobar, 
up to 390K for the p=65000bar isobar and up to 380K for the p=60000bar isobar. 
At higher temperatures ice VIII also becomes a plastic crystal phase. Thus ice VIII becomes
mechanically unstable (transforming into a plastic crystal phase) at temperatures similar to those
found for ice VII.  The mechanical stability limit of ice VIII (where it transforms into a 
plastic crystal phase) seems to be about 10K higher than that of ice VII.
We did not observe any spontaneous transformation 
of ice VIII into ice VII when heated . This is not to say that this transition does not exist from 
a thermodynamic point of view (in fact it will be shown later that this transition indeed 
exists) but rather that the system is not able to overcome the 
free energy nucleation barrier separating ice VIII from ice VII within the length of our 
simulations. 
 In any case for pressures below 70000 bar it is not possible to have neither 
ice VIII nor ice VII as mechanically stable phases for temperatures above 390K.
This provides further evidence that the plastic crystal phase obtained spontaneously from 
the liquid at 440K must indeed be thermodynamically stable at least at  
high temperatures.  In any case, the definite proof of that 
will be provided by the free energy calculations. 
   In Fig. \ref{fdr_vii_viii}
the radial distribution functions 
of ice VIII are compared to those of ice VII at 300K and 70000 bar. 
They are clearly different showing that both solids  are mechanically stable 
(and clearly distinguishable) under these thermodynamic conditions.

\begin{figure}[!h]
\begin{center}
\caption{Oxygen-oxygen, oxygen-hydrogen and hydrogen-hydrogen
radial distribution functions for ice VII at 300K and 
70000 bar (solid line) and for
the plastic crystal phase at 400K and 70000bar (dashed-dotted line).}
\label{fdr_vii_plastic} 
\includegraphics[clip,height=0.35\textheight,width=0.60\textwidth,angle=-0]{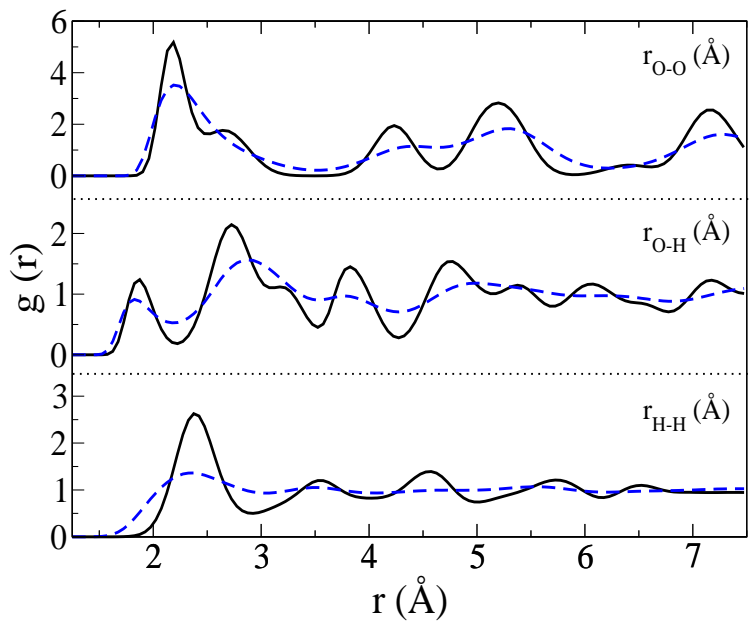} 
\end{center}
\end{figure}

  Let us now present the structural properties for the plastic crystal phase.  
In Fig. \ref{fdr_vii_plastic}, 
the radial distribution functions for the plastic crystal phase 
at 400K and 70000 bar are compared to those
of ice VII at 300K and 70000bar for the O-O, O-H and H-H correlation functions. 
 As it can be seen from the oxygen-oxygen correlation 
function, the peaks of the O-O distribution in the plastic crystal phase correspond to those of ice VII, 
although the higher temperature and lower density of the plastic crystal solid provokes peaks
with a more diffuse character. 
Changes in the $g_{O-H}$ distribution function are significant, but the most important change occurs 
in the $g_{H-H}$ correlation function. It has a liquid like aspect in the plastic crystal solid whereas its long range order is clearly visible in ice VII. Thus the $g_{H-H}$ distribution function is quite 
useful to identify the existence of a plastic crystal phase. 

 We have also determined the probability distribution of the polar 
angles ($\theta$ and $\phi$ of the OH bonds). The x axis is located on the {\bf a} vector of the unit cell, the y axis is located on the {\bf b} vector of the unit cell, and the z axis is located along 
the {\bf c} vector of the unit cell. 
The distribution function $f(\theta)$ is defined as :

\begin{equation}
 f(\theta) =    \frac{ N(\theta)}{ 2 N \Delta \theta }
\end{equation}
where $N(\theta)$ denotes the number of $OH$ bonds with polar angle between $\theta$ and 
$\theta+ \Delta \theta$  and  2N is the number of OH bonds (i.e twice the number of molecules N).
The distribution function $f(\phi)$ is defined as :
\begin{equation}
 f(\phi) =    \frac{ N(\phi)}{ 2 N \Delta \phi}
\end{equation}
 In Fig. \ref{figure_ftheta_vii_viii} the 
functions  $f(\theta)$ and $f(\phi)$ are presented for ices VIII and VII at 
300K and 70000bar. As it can be seen the peaks of $f(\theta)$ are located at $54.74$ and 
$(180-54.74)$ where 54.74 is the angle in ice VII between one of the diagonals of the cube and a line connecting the
center of two opposite faces. The four peaks in $f(\phi)$ are separated by 90 degrees as it should
be. In Fig. \ref{figure_ftheta_vii_viii} 
 the functions  $f(\theta)$ and $f(\phi)$ are also presented for the plastic 
crystal phase at 400K and 70000bar. As it can be seen both distributions present less 
structure than the one of ices VII and VIII. It is also clear that even in the plastic 
crystal phase the OH vectors prefer to point out to the contiguous oxygens atoms. Therefore 
the angular distribution is not uniform even in the plastic crystal phase 
(i.e the OH vector does not distribute uniformly on the surface of a sphere). The molecules, although
able to rotate prefer to point out the OH vectors to any of the 8 O atoms forming the 
first coordination layer. However, we checked that as the temperature increases the 
$f(\theta)$ and $f(\phi)$ distributions become somewhat more uniform. In summary the molecules 
rotate almost freely in the plastic crystal phase, but the OH vectors prefer to point out to the
contiguous oxygen atoms. 
An additional interesting comment is that for ices VIII and 
VII at 300K and 70000 bar and for the length of the
simulations performed in this work , a certain individual molecule presents a fixed value 
of $\theta$ and $\phi$ (except for a little bit of thermal vibration). However for the 
plastic crystal phase each individual molecule jumps quite often from one of the 
peaks of the distribution to another peak. These flipping or jumping moves occur quite 
often for each molecule within the length of the simulation runs considered here (at 400K in average
each flipping requires about 8000 Monte Carlo cycles). 
The ice VIII to ice VII transition represents the change from proton order to 
proton disorder within the two sub lattices of the solid, whereas the ice VII to 
plastic crystal phase transition represents the disappearance of the two sublattices.

  Let us now change to the determination of coexistence points. 
We have determined the free energy for ices VIII, VII and for the 
plastic crystal phase (Table \ref{free_energy_table}).
The free energy calculation was performed for ice VIII at 300K and for a density which 
corresponds to that of the system at 70000 bar. The free energy of ice VII was computed
at 300K and for a density which corresponds to that of 70000bar. Finally the free energy 
of the plastic crystal phase was computed at 400K and for a density which corresponds 
to that of the system at 70000 bar. Details about the free energy calculations in water 
solid phases have been provided elsewhere\cite{vega_review}. Let just mention the procedure used to 
compute the free energy of the plastic crystal phase as it differs from that used
for the rest of solid phases of water.  Firstly translational springs  
connecting the oxygen atoms to the lattice positions of the bcc 
structure were switched on (in the plastic
crystal phase the oxygens form a bcc solid). After that, and while keeping the translational 
springs, the charges of the system were switched off ending with a system of LJ atoms connected 
by springs to a bcc solid structure. Free energies changes can be computed easily along this
path in which the Hamiltonian of the system is changed. Notice that no phase transition 
occurs along the
integration since the system is in a plastic crystal phase along the entire integration path. 
Finally the free 
energy difference between a LJ particle connected with translational springs to a bcc lattice, 
and an ideal Einstein crystal (with springs connecting to the bcc lattice but without 
intermolecular interactions) is computed by a perturbative approach. This integration path  
is inspired by a recent article by Lindberg and Wang  
in which the dielectric constant of ice Ih was 
computed by switching off the charges of the molecules while having translational springs\cite{dielectric_constant_ice}.  
In Table III the free energies for the considered thermodynamic states are 
reported. Let us just mention, that the Pauling degeneracy entropy was added to the 
free energy of ice VII. This contribution was not added neither to ice VIII nor to the 
plastic crystal phase.

\begin{figure}[!h]
\begin{center}
\caption{Evolution of the density with time as obtained from MD simulations.
All results were obtained for T=400K.
Lines from the top to the bottom correspond to the pressures 70000,65000,58000 
and 55000bar respectively.}
\label{figure_coexistencia}
\includegraphics[clip,height=0.35\textheight,width=0.60\textwidth,angle=-0]{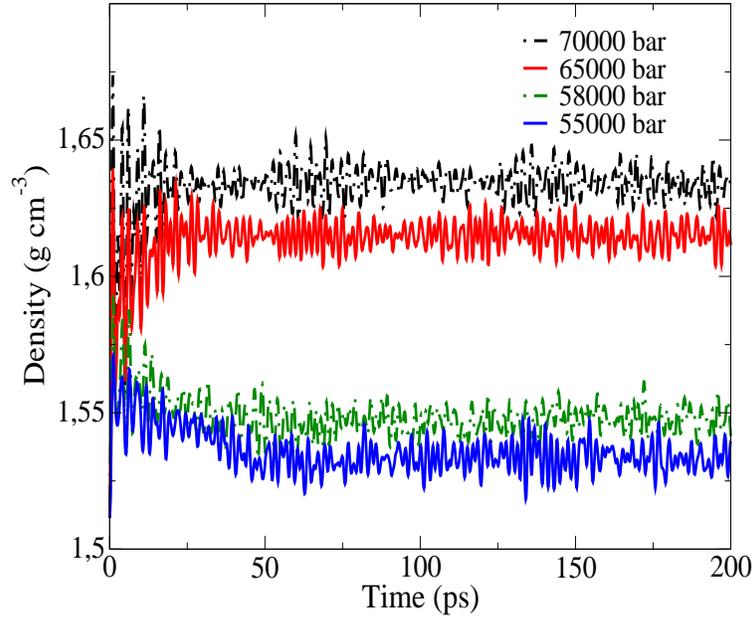}
\end{center}
\end{figure}

 As it can be seen the Gibbs free energy of ice VII is smaller than that of ice VIII. 
Both phases have similar densities and internal energies and the higher stability of ice VII 
is mainly due to the contribution of the Pauling entropy (which is not present in ice VIII).
Therefore
for T=300K and p=70000bar ice VII is more stable than ice VIII. By using thermodynamic integration the ice VII to ice VIII transition at 70000 bar has been located at T=69K. 
In the same way the VII to plastic crystal transition  
has been located to occur at 377K for a pressure of 70000 bar. This value is consistent with the hysteresis
loop found at 70000bar between the temperatures 350 and 390. Finally by using thermodynamic 
integration the fluid-plastic crystal transition was found to occur at a pressure of  
$p=62000 bar$ for the 400K isotherm. 
The fluid-plastic crystal transition was also determined by using 
direct coexistence simulations. An equilibrated configuration of the 
 plastic crystal phase (432 molecules) was located on the left hand side of a simulation box, 
and put into contact with an equilibrated configuration of the fluid having 432 molecules. 
Molecules of the fluid phase overlapping with those of the solid phase at the fluid-solid 
interface were deleted. Thus the total number of molecules in the direct coexistence simulations was
of 858. We then performed MD simulations using Gromacs, while keeping the temperature at 
400K. Runs at different pressures were performed. At pressures below the melting point 
the solid phase will melt, whereas at pressures higher than the coexistence pressure the 
fluid phase will crystallise. The direct coexistence technique, was pioneered by 
Ladd and Woodcock\cite{woodcock1,woodcock2,woodcock3}, and used recently by several 
authors\cite{broughton5,laird98,morris02,thompson_nitromethane_defects_and_twophases,karim88,bryk02,ramon06,abascal06,kroes_hydrate,wang05,kusalik_hydrate,noya_hs_interface}.
In Fig. \ref{figure_coexistencia}, the evolution of the density  
with time is shown for several pressures.
As it can be seen for the low pressures  (55000 and 58000 bar) the density 
decreases quickly indicating the
melting of the plastic crystal. For high pressures (i.e 65000, 70000 bar) the density 
of the system increases quickly indicating the freezing of the fluid into a plastic 
crystal phase. Thus the value of 62000 bar obtained from free energy calculations 
is consistent with the results obtained from direct coexistence simulations\cite{ramon06}. 
Notice that in the direct coexistence simulations the melting of the plastic crystal, or the freezing of the water occurs
in about 0.2ns, which is about two orders of magnitude smaller than the time 
required to determine accurately the melting point of ice Ih at room pressure by 
direct coexistence simulations\cite{ramon06}. This is probably due to two different facts. 
First, the temperature is high (400K instead of the temperature of about 250K used 
in the ice Ih-water direct interface simulations). Secondly the growth of the crystal
is a somewhat more complex process for ice Ih since each water molecule must be located
in a certain position and with a certain orientation to allow the ice Ih to grow. However, 
in the plastic crystal phase, the molecule must find the location, but the orientation 
is not so important since the molecules are rotating in the plastic crystal phase anyway. 

\begin{figure}[!h]
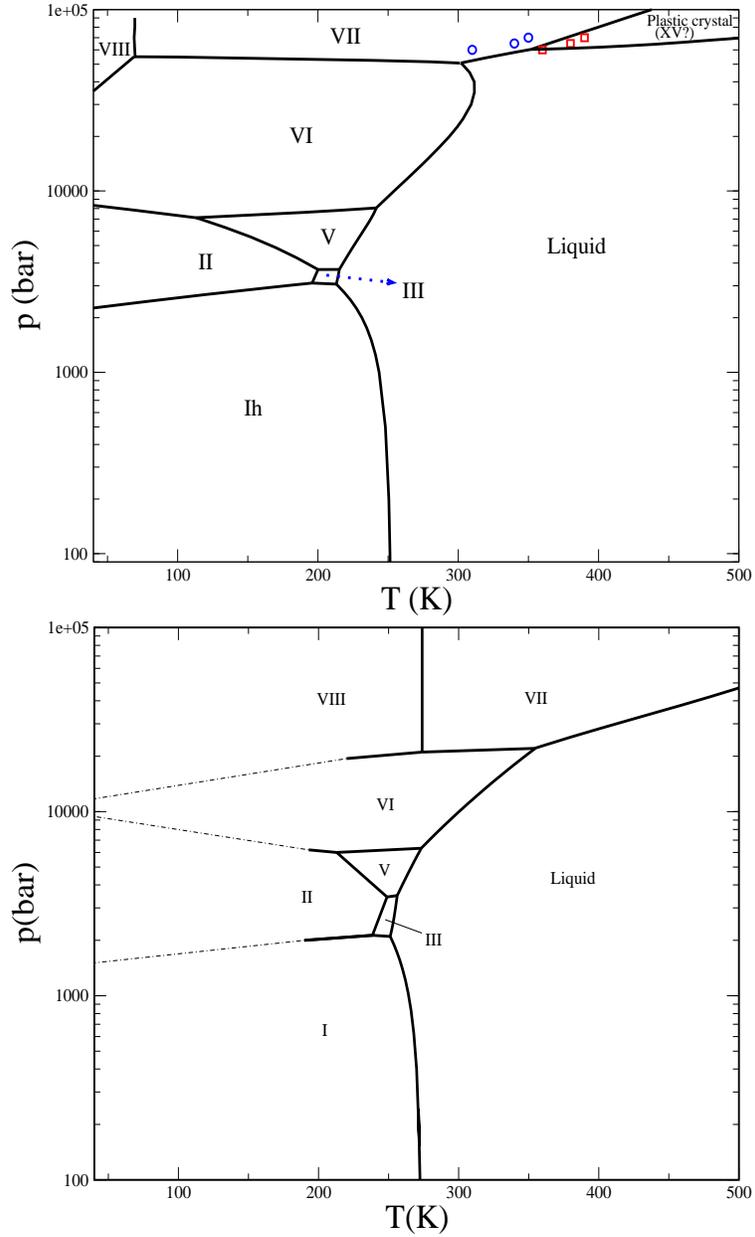

\begin{center}
\caption{a) Global phase diagram for the TIP4P/2005 model. The squares correspond 
to the stability limit of ice VII when heated, whereas the circles correspond to 
the stability limit of the plastic crystal phase when cooled. b) Experimental 
phase diagram of water.}
\label{figure_phase_diagram_high_pressures}
\includegraphics[clip,height=0.35\textheight,width=0.60\textwidth,angle=-0]{phase_diagram_high_pressures.eps}
\includegraphics[clip,height=0.35\textheight,width=0.6\textwidth,angle=-0]{diagramaexpagua.eps}
\end{center}
\end{figure}

 Once  an initial coexistence point has been found for the  VIII-VII,  VII-plastic 
and VII-fluid transitions, the rest of the coexistence lines will be obtained by 
Gibbs Duhem simulations. In Fig. \ref{figure_phase_diagram_high_pressures}
these three coexistence lines are plotted. 
As it can be seen the fluid-plastic crystal and the VII-plastic crystal coexistence 
lines met at a triple point located around 352K and 60375 bar. This triple point 
can be used as initial coexistence point of the  VII-fluid coexistence line. 
The fluid-VII coexistence line obtained from Gibbs Duhem simulations has also 
been plotted in Fig. \ref{figure_phase_diagram_high_pressures}. At 300K 
the Gibbs Duhem integration predicts a fluid-VII coexistence 
pressure of about 51000 bar, which is consistent with the coexistence pressure predicted 
from free energy calculations. The melting curve of ice VI intersects the melting curve of 
ice VII, generating a fluid-VI-VII triple point (the precise location is 
301K and 50742 bar). This triple point can be used as
initial point for the VI-VII coexistence curve. The VI-VII coexistence curve 
intersects the VIII-VII coexistence curve at about 55000 bar and 69K. This generates 
an VI-VII-VIII triple point which can be used to initiate the VI-VIII 
coexistence line. The global phase diagram of the TIP4P/2005 (including the high pressure 
phases) is presented in Fig. \ref{figure_phase_diagram_high_pressures}.  The experimental phase 
diagram is also presented in Fig. \ref{figure_phase_diagram_high_pressures}. 
The coexistence lines are given in tabular form in Tables IV-VI.
The triple points are given in Table \ref{triple_points}.

\begin{figure}[!h]
\begin{center}
\caption{Equation of state of ices VII (open circles) and VIII (open squares) for 
T=300K as obtained from computer simulation for the TIP4P/2005 model. The solid line 
on the right hand side correspond to the experimental results as given by
Fei et al.\cite{JCP_1993_99_5369}.}
\label{figure_eos_300K}
\includegraphics[clip,height=0.35\textheight,width=0.60\textwidth,angle=-0]{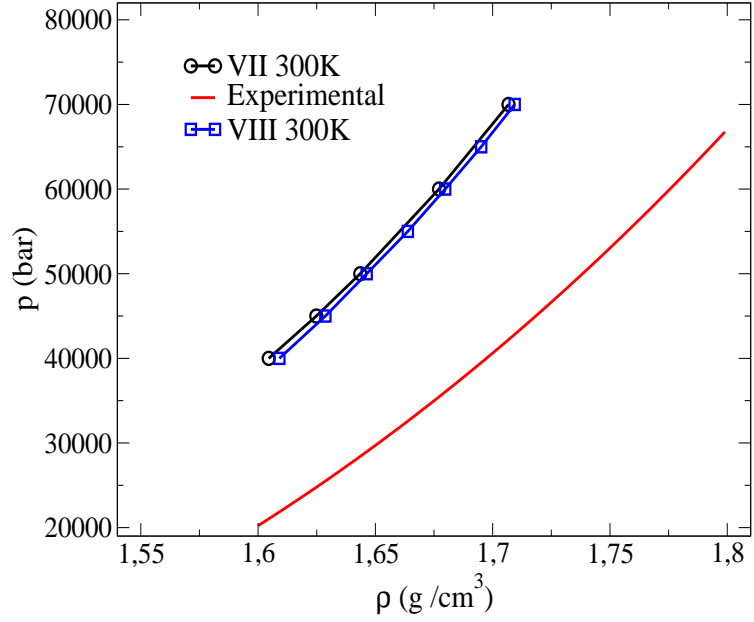}
\end{center}
\end{figure}

As it can be seen the TIP4P/2005 reproduces qualitatively the phase diagram of 
water. Ice VIII is stable only at low temperatures. The coexistence line between 
ice VIII and ice VII is almost a vertical line. This is due to the fact that the densities 
of ices VII and VIII are practically identical for a certain temperature and pressure. 
There is a certain regime of temperatures where ice VII coexists with the liquid phase.
There are three  significant qualitative differences between the phase diagram of 
TIP4P/2005 and the experimental one. The first is that the VIII-VII transition temperature seems
to be low as compared to experiment. It seems that models with a TIP4P geometry tend to underestimate
the temperature of the order-disorder transitions (between ices having the same arrangement of oxygens but  differing in the proton ordering). In fact we found in previous work\cite{vegautrecht} 
that also the XI-Ih transition was predicted to occur for the TIP4P model at a temperature lower than 
found experimentally.  
 The second  is the  appearance of a plastic crystal phase for
the TIP4P/2005 model. This plastic crystal phase has not been reported so far for 
real water. The third is the appearance of re-entrant melting in the melting curve of 
ice VI which is not found in the experimental phase diagram. The reason for the 
appearance of re-entrant melting in the melting curve of ices (from ice Ih up to ice VI) 
has been discussed previously \cite{sanz1}. The fluid-VI-VII triple point is located at 
301K and 50742 bar, to be compared with the experimental value which is 
located at a temperature of 355 K and a pressure of about 22000 bar. Thus, it seems that
to bring the phase diagram of TIP4P/2005 into closer agreement with experiment the 
stability of ice VII should increase. In this way the fluid-VI-VII triple point 
would occur at lower pressures and the re-entrant portion of the melting curve of ice 
VI would occur in the metastable part of the melting curve. In Fig. \ref{figure_eos_300K} the 
equation of state of ices VIII and VII at 300K are plotted as a function of pressure. 
For pressures below 35000 bar ice VIII (as described by TIP4P/2005) is not mechanically 
stable and melts into a liquid
phase. For pressures below 40000 bar ice VII (as described by TIP4P/2005) is not 
mechanically stable and melts into a 
liquid phase. As it can be seen for a certain temperature and pressure ice VIII yields 
a slightly higher density than ice VII (the densities are quite similar anyway as also pointed
out by Bartok and Baranyai\cite{bartok_baranyai}).  
As it can be seen the TIP4P/2005 model is not able to describe accurately 
the experimental values of the densities of ice VII. In fact it tends to underestimate 
the experimental values of the densities by about 5 per cent. This is striking since the 
TIP4P/2005 is able to describe the densities of ices Ih, II, III, IV, V, VI, IX and XII (for 
pressures up to  20000 bar ) within an error of about 1 per cent and for liquid water up to pressures of about 40000 bar with an error of about 1 per cent. 
Thus it seems that 
there is something wrong in the model which prevents of describing accurately the densities of ices VIII and VII at high pressures. 
 Notice that other water models as SPC, SPC/E and TIP4P do also 
underestimate significantly the density of ice VII at high pressures. 
 Further work is needed to understand the 
origin of that. 
One may suggest that the absence of polarisability, or the use of a non very accurate 
description of the repulsive part of the potential (by the LJ potential), or even the 
absence of LJ sites located on the hydrogen atoms (to introduce a further degree of 
anisotropy) may be responsible.  
 The only model that reproduces  the experimental  
densities of ice VII is TIP5P. This is shown in Table \ref{density_ices}. 
However, this is not for free since this model overestimates 
by about 5-8 per cent\cite{pccpgdr} the densities of ices Ih, II, III, IV, V, VI, IX and XII. Thus TIP5P 
reproduces the densities of ice VII, but fails completely in describing the densities of 
the other polymorphs of water
\cite{JCP_1993_99_5369,bookPhysIce,fortes_jac05}.

  Let us finish by presenting the value of the dielectric constant of the 
plastic crystal phase. 
  It should be pointed out that standard
  computer simulations (MD or standard MC) can not be used to determine the dielectric constant of
  ices. Typical lengths of the simulations (several ns or 
hundred of thousands
   of cycles) are not sufficient 
to sample the fluctuations of the polarisation of the solid.
   To compute dielectric constants of  ices it is necessary either to introduce special
  moves within the Monte Carlo program, as those proposed by Rick and 
  Haymet \cite{JCP_2003_118_09291,rick05} or to use special
  techniques as that proposed recently by Lindberg and  Wang\cite{dielectric_constant_ice}.  
  Thus from our study we could not report
  the dielectric constant of ices VIII and VII. However we could compute the dielectric constant 
  for the plastic crystal since the molecules are able to rotate rather quickly 
  provoking important fluctuations of the total polarisation.
  We obtained a value of about 96(10) for the plastic crystal phase at 400K and 
  70000 bar. The dielectric constant of the plastic crystal 
  phase
  (assuming it exists in real water) has not been reported. The experimental 
  value\cite{johari_whalley} of
  the dielectric constant of ice VII  at room temperature and for a pressure
  of 23300 bar is of 105.
  Although it is difficult to compare dielectric constants obtained for somewhat different
  structures (ice VII and the plastic crystal phase are not the same solid), and somewhat
  different thermodynamic conditions, the comparison between both values appears as
  reasonable. For liquid water at room temperature and pressure the TIP4P/2005 
  predicts a dielectric constant
  of about 60, whereas the experimental value is of 78. Thus the model underestimates the
  dielectric constant of liquid water by about 20 per cent. 
  Also for ice Ih, TIP4P models
  seriously underestimate the dielectric 
  constant\cite{dielectric_constant_ice,JCP_2003_118_09291,rick05}. 
  Assuming that similar behaviour 
  could be found for liquid water and the plastic ice,
  the prediction of the model for the plastic crystal phase 
  seems to be lower than the experimental value of ice VII.
  The absence of polarisability is
  likely to be responsible for this deviation.

  \section{ Conclusions }
  In this work the high pressure region of the phase diagram of water has been determined
  by computer simulation by using the TIP4P/2005 model. By performing free energy calculations
  the relative stability of ices VIII and VII, and the melting curve of ice VII was obtained.
  When compressing water at 440K, the freezing of water into a solid phase occurred in a few
  ns. The analysis of the solid allows to conclude that 
  it is formed by a body center cubic lattice of oxygens, with
  the molecules rotating significantly. In summary a plastic crystal phase.
  It has also been found that ices  VIII and VII transform into a plastic crystal phases
  at temperatures above 390K and pressures below 70000 bar.  By performing 
  free energy calculations
  the melting curve of the plastic crystal phase and the coexistence lines between the plastic
  crystal phase and ice VII could be determined. In this way a rather extensive region
  of the phase diagram for
  the TIP4P/2005 could be reported by the first time. The possible existence of other solid phases or even of other plastic crystal phases for the model should not be discarded yet. 
  The TIP4P/2005 is able to describe
  qualitatively the phase diagram of water.  The model (in common with the popular
  SPC/E and TIP4P models) fails to describe the density of ice VII along the room temperature 
  isotherm, and underestimates with respect to experiment the stability of ice VII. This
  is in contrast with the excellent description provided for other coexistence lines and for
  the densities of other polymorphs appearing at pressures below 20000 bar.
  Thus the model does not allow  to describe quantitatively the
  behaviour of water at high pressures. 
  Further work is needed to understand the origin of that.  On the other hand the model
  predicts the existence of a new phase, a plastic crystal phase. Notice that all transitions
  found between the plastic crystal phase and other phases (liquid or ice VII) were found to
  be first order, so that if the plastic crystal phase exists in real water there should
  be a first order phase transition separating it from ice VII. Interestly Takii, Koga and Tanaka
  suggested that the two set of melting curves reported experimentally at 100000 bar could be due
  to the fact that these two groups were indeed measuring two different transitions, namely
  the ice VII to plastic crystal (that would correspond to the temperature of  600K given by 
  Mishima and Endo\cite{mishima78}) and the plastic crystal to liquid phase (that would correspond
  to the temperature of 687K given by Datchi, Loubeyre and LeToullec \cite{frenchgroup} ). 
  Further experimental work is needed to clarify this point. 

    It would be of interest to study if such a plastic crystal phase does 
    indeed exist for real water. If so it would corresponds to 
   ice XV (the precise roman numeral may change if other solid phases of water are found experimentally before ), and would probably be one of the  few cases where computer
  simulation anticipates an experimental result.
  In case it exists it may dominate the melting curve 
  of water at high temperatures.  
 The nearly spherical shape
 of water makes possible the formation of a plastic crystal phase at high temperatures 
where the strength of the hydrogen bond in $k_{B}T$ units is rather small and where the molecular
shape (i.e repulsive forces) play the leading role in the process of freezing. For hard 
dumbbells or spherocylinders the maximum anisotropy (as given by the ratio between the bond length and 
the width of the molecule) which allows the formation of 
a plastic crystal phase is of about $0.4$ \cite{vega92b,vega92c,MP_1992_77_0803_photocopy,JCP_1997_107_02696,JCP_1997_106_00666,ACP_2000_115_0113_nolotengo}. 
Molecules with  higher anisotropy  freeze into an orientally ordered solid. 
Thus the formation of plastic crystal phases requires moderate anisotropy in the repulsive
part of the potential (the importance of the attractive part decreases substantially at high 
temperatures). This seems to be the case of water, at least when described as a LJ center plus 
charges. Besides the experimental importance of finding a plastic crystal phase for
water, its presence or absence should also have an impact on the community performing
computer simulations of water. Its presence would indicate that current water models
are able to predict qualitatively this new feature of the phase diagram of water. 
Its absence would also
be significant since it would indicate that current water models are too spherical and
they should be modified so as to predict the disappearance of the plastic crystal phase. 

\begin{figure}[!h]
\begin{center}
\caption{Snapshots of ices. a) Snapshot of ice VIII at 300K and 70000bar. b) Snapshot of ice VII 
at 300K and 70000bar. c) Snapshot of the plastic crystal phase at 400K and 70000 bar.}
\label{figure_snapshots}
\includegraphics[clip,height=0.35\textheight,width=0.60\textwidth,angle=-0]{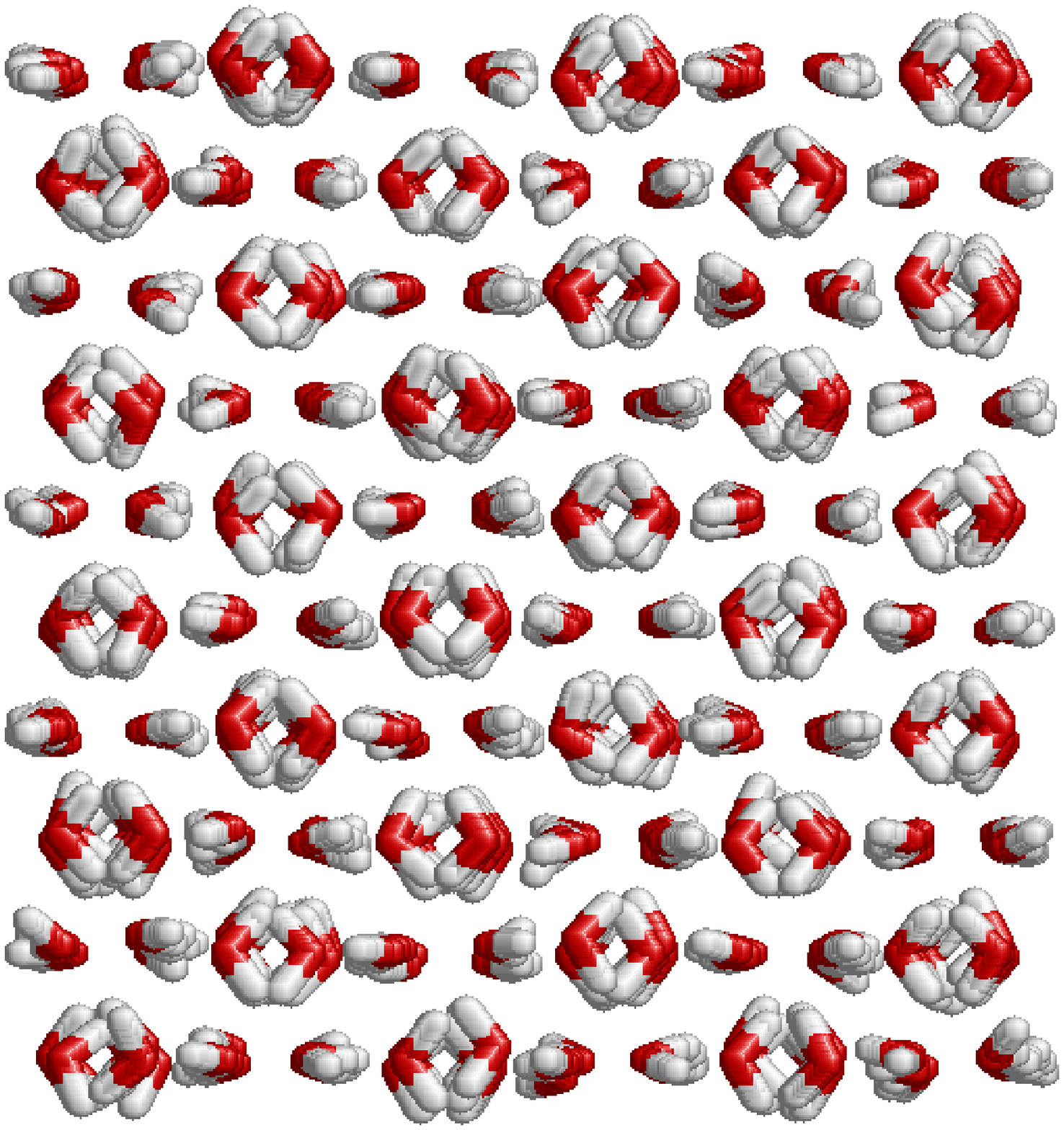}
\includegraphics[clip,height=0.35\textheight,width=0.59\textwidth,angle=-0]{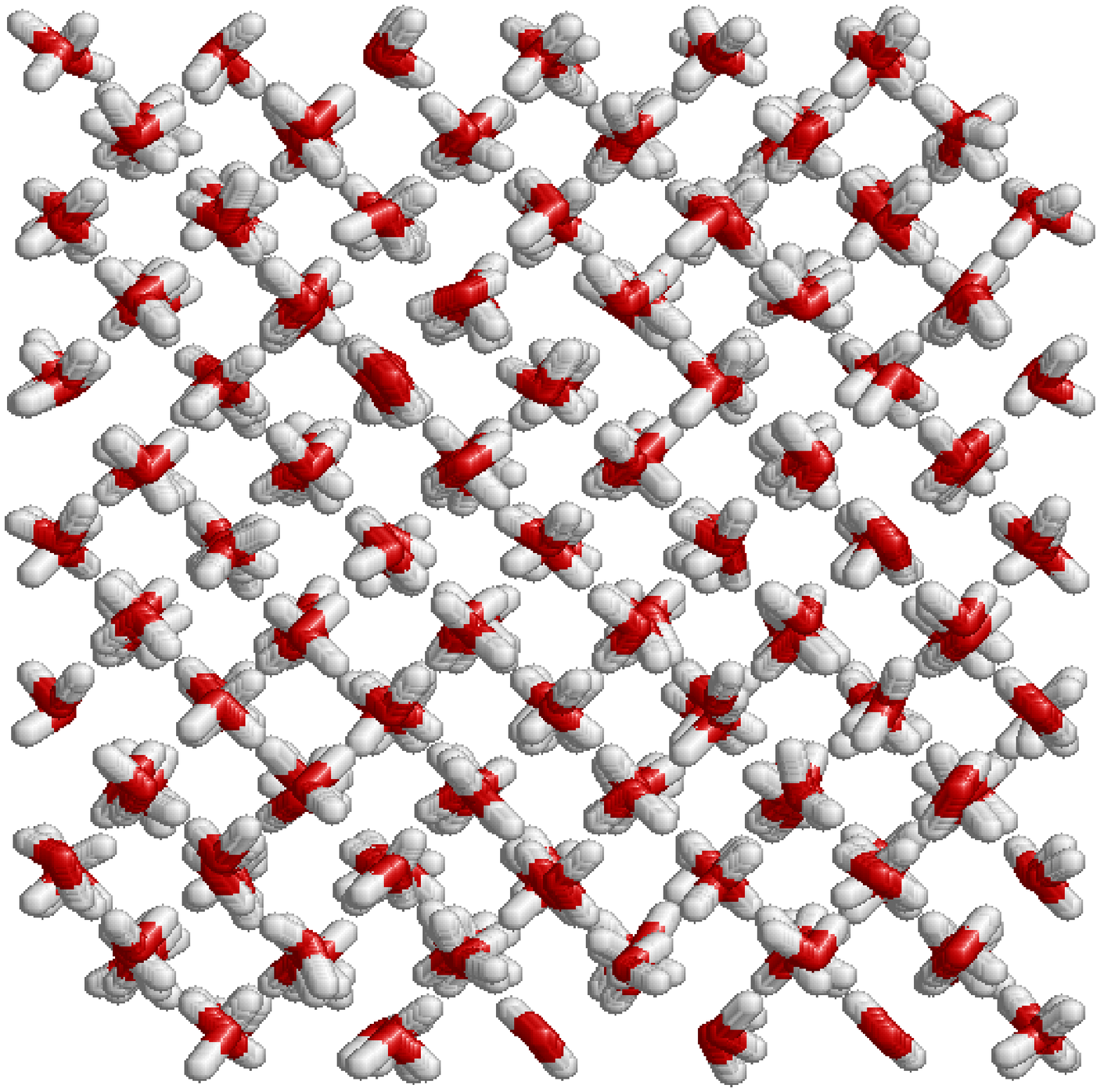}
\includegraphics[clip,height=0.35\textheight,width=0.62\textwidth,angle=-0]{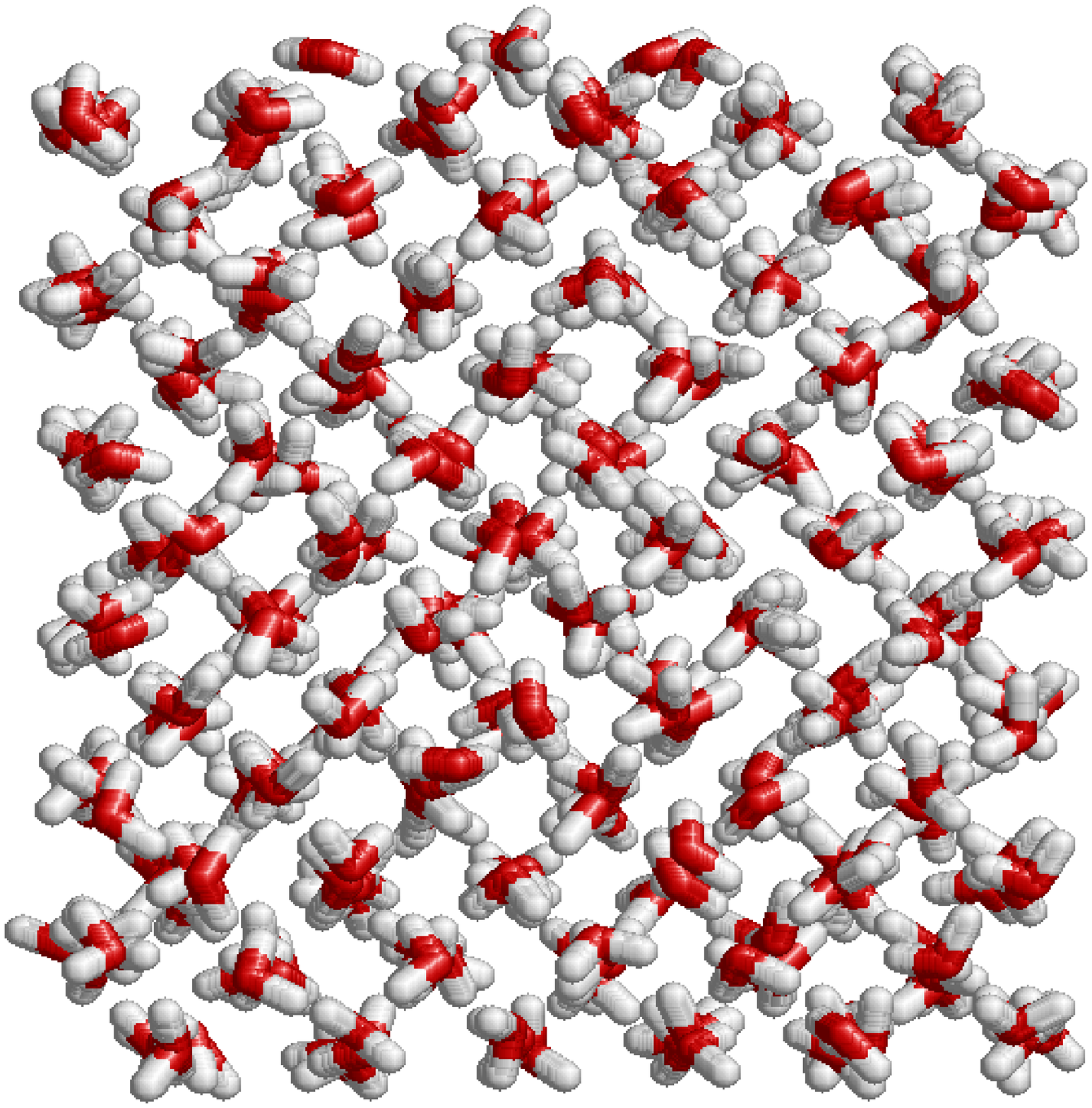}
\end{center}
\end{figure}

    Since a picture is worth a thousand words let us finish by presenting an instantaneous 
snapshot of ices VII, VIII and of the plastic crystal. This is done in Fig. \ref{figure_snapshots}.
The orientational disorder of the plastic crystal phase is shown clearly in the figure. 

\section{acknowledgements} 
This work was funded by grants FIS2007-66079-C02-01
from the DGI (Spain), S-0505/ESP/0229 from the CAM, 
and 910570 from the UCM.
M.M.Conde would like to thank Universidad Complutense by the award of a
PhD grant. J. L. Aragones would like to thank the MEC by the award of a pre-doctoral
grant.  
E.G.N. wishes to thank
the Ministerio de Educaci\'on y Ciencia
and the Universidad Complutense de Madrid for a Juan de la Cierva fellowship.
Helpful discussions with
L.G.MacDowell and J.L.F.Abascal are gratefully acknowledged.

\bibliographystyle{./apsrev}

\newpage

\begin{table}
\caption{ Parameters of the TIP4P/2005 model. The distance between the oxygen and hydrogen sites 
is $d_{OH}$. The angle formed by hydrogen, oxygen, and the other hydrogen atom is 
denoted by H-O-H. The LJ site is located on the oxygen with parameters $\sigma$ and $\epsilon/K$. 
The charge on the proton is $q_{H}$. The negative charge is placed in a point $M$ at a distance 
$d_{OM}$ from the oxygen along the H-O-H bisector.}
\label{tip4p_2005_parameters}
\begin{tabular}{cccccc}
\hline
$d_{OH}$ (~\AA) &\hspace{0.2cm} H-O-H &\hspace{0.2cm} $\sigma$(~\AA)&\hspace{0.2cm} $\epsilon/k_{B}$(K)& \hspace{0.2cm} $q_{H}$(e)& \hspace{0.2cm}$d_{OM}$(~\AA) \\
\hline
0.9572 & \hspace{0.2cm}104.52 & \hspace{0.2cm}3.1589 & \hspace{0.2cm}93.2 &\hspace{0.2cm} 0.5564 &\hspace{0.2cm} 0.1546\\
\end{tabular}
\end{table}

\newpage
\clearpage

\begin{table}
  \caption{ Stability limits of ice VII when heated and of the plastic crystal phase when 
cooled as obtained from NpT simulations at three different pressures. }
      \label{hysteresis_lopp_table}
      \begin{tabular}{ccc}
      \hline
p/bar  &  \hspace{0.3cm}Ice VII-plastic crystal {\it (heating)}& \hspace{0.3cm} Plastic crystal-ice VII {\it (cooling)}   \\ 
\hline
70000    &    390        &                     350 \\ 
65000    &    380        &                     340  \\
60000    &    360        &                      310 \\
\end{tabular}
\end{table}
    
\clearpage
\newpage  
\begin{table}
  \caption{ Helmholtz free energy for ices VIII, VII and for the plastic crystal phase.
 The Gibbs free energy was computed after adding pV to the Helmholtz free energy. 
The data marked with an asterisk correspond to calculations using the structure ice VII 
as obtained from the cooling of the 
plastic crystal.
The density 
is given in $g/cm^3$.}
      \label{free_energy_table}
      \begin{tabular}{cccccc}
      \hline
Phase  & \hspace{0.2cm}T/K  & \hspace{0.2cm} p/bar    & \hspace{0.2cm} $\rho$   & \hspace{0.2cm}A/Nk$_B$T & \hspace{0.2cm} G/Nk$_B$T   \\
\hline
Ice VIII   & \hspace{0.2cm}   300    &\hspace{0.2cm} 70000 &\hspace{0.2cm} 1.709 &\hspace{0.2cm}-9.36 &\hspace{0.2cm} 20.23   \\
Ice VII    &  \hspace{0.2cm}  300    &\hspace{0.2cm} 70000 & \hspace{0.2cm}1.707 &\hspace{0.2cm}-9.75 & \hspace{0.2cm}19.87   \\
Ice VII$^{*}$    &  \hspace{0.2cm}  300    &\hspace{0.2cm} 70000 & \hspace{0.2cm}1.707 &\hspace{0.2cm}-9.74 & \hspace{0.2cm}19.88   \\
Plastic    &   \hspace{0.2cm} 400    &\hspace{0.2cm} 70000 & \hspace{0.2cm}1.638 &\hspace{0.2cm}-6.46 & \hspace{0.2cm}16.69   \\
Water      &  \hspace{0.2cm}  300    &  \hspace{0.2cm} 1292    & \hspace{0.2cm}1.050 &\hspace{0.2cm}-15.45 & \hspace{0.2cm}-14.56  \\
\end{tabular}
\end{table}

\clearpage
\newpage
\begin{table}
  \caption{  Melting curve of the plastic crystal phase of the TIP4P/2005 
as obtained from free energy calculations and Gibbs Duhem integration (with isotropic NpT simulation performed for the fluid and for the cubic plastic crystal phase). The densities are given in $g/cm^{3}$. The residual internal energies are given in  
in kcal/mol.}
      \label{coexistence_lines_table_1}
      \begin{tabular}{cccccc}
      \hline
  T/K  &  p/bar    & $U_{1}$  &  $U_{2}$  & $\rho_1$   & $\rho_2$   \\
      \hline
 {\bf  fluid-plastic crystal} \\
  340.00  &   60193  &  -10.45  &  -10.03  &   1.574  &   1.622  \\
  360.00  &   60486  &  -10.24  &   -9.85  &   1.572  &   1.613  \\
  380.00  &   61135  &  -10.04  &   -9.73  &   1.565  &   1.610  \\
  400.00  &   62000  &   -9.87  &   -9.58  &   1.562  &   1.609  \\
  420.00  &   63108  &   -9.75  &   -9.43  &   1.565  &   1.607  \\
  440.00  &   64390  &   -9.52  &   -9.30  &   1.564  &   1.608  \\
  460.00  &   66042  &   -9.35  &   -9.12  &   1.560  &   1.607  \\

\end{tabular}
\end{table}

\newpage

\clearpage
\begin{table}
  \caption{  Melting curves of ices VI and VII for the TIP4P/2005 model as obtained from free energy calculations and Gibbs Duhem integration. The densities are given in $g/cm^{3}$. The internal energies 
in kcal/mol (only the residual part of the internal energy is reported).}
      \label{coexistence_lines_table_2}
      \begin{tabular}{cccccc}
      \hline
  T/K  &  p/bar    & $U_{1}$  &  $U_{2}$  & $\rho_1$   & $\rho_2$   \\
      \hline
 {\bf fluid-VI}\\
  244.97  &    8540  &  -12.23  &  -12.96  &   1.245  &   1.359  \\
  254.39  &   10000  &  -12.02  &  -12.91  &   1.265  &   1.365  \\
  265.14  &   12000  &  -11.98  &  -12.84  &   1.286  &   1.373  \\
  278.27  &   15000  &  -11.77  &  -12.76  &   1.316  &   1.385  \\
  293.67  &   20000  &  -11.55  &  -12.64  &   1.365  &   1.406  \\
  303.39  &   25000  &  -11.38  &  -12.55  &   1.396  &   1.427  \\
  309.21  &   30000  &  -11.26  &  -12.47  &   1.435  &   1.447  \\
  311.52  &   35000  &  -11.21  &  -12.39  &   1.464  &   1.467  \\
  311.39  &   40000  &  -11.15  &  -12.32  &   1.492  &   1.485  \\
  308.14  &   45000  &  -11.03  &  -12.26  &   1.518  &   1.503  \\
  302.87  &   50000  &  -10.98  &  -12.20  &   1.542  &   1.520  \\
  293.92  &   55000  &  -10.97  &  -12.15  &   1.572  &   1.537  \\
  281.74  &   60000  &  -10.89  &  -12.11  &   1.596  &   1.553  \\

 {\bf fluid-VII} \\
  352.00  &   60375  &  -10.30  &  -10.46  &   1.578  &   1.663  \\
  340.00  &   58066  &  -10.42  &  -10.57  &   1.570  &   1.658  \\
  320.00  &   54276  &  -10.74  &  -10.74  &   1.565  &   1.651  \\
  300.00  &   50988  &  -10.90  &  -10.90  &   1.549  &   1.646  \\

\end{tabular}
\end{table}

\clearpage

\newpage
\begin{table}
  \caption{ Solid-solid coexistence lines of the TIP4P/2005 model for the high pressure polymorphs
(ice VII,VIII and plastic crystal) as obtained from free energy calculations and Gibbs Duhem integration. The densities are given in $g/cm^{3}$. The internal energies 
in kcal/mol (only the residual part of the internal energy is reported).}
      \label{coexistence_lines_table_3}
      \begin{tabular}{cccccc}
      \hline
  T/K  &  p/bar    & $U_{1}$  &  $U_{2}$  & $\rho_1$   & $\rho_2$   \\
      \hline
 {\bf VII-plastic crystal} \\
  440.00  &  101354  &   -8.99  &   -8.49  &   1.749  &   1.728  \\
  420.00  &   90168  &   -9.44  &   -8.81  &   1.729  &   1.698  \\
  400.00  &   80146  &   -9.76  &   -9.19  &   1.707  &   1.674  \\
  390.00  &   75596  &   -9.82  &   -9.33  &   1.693  &   1.659  \\
  377.00  &   70000  &  -10.04  &   -9.53  &   1.681  &   1.645  \\
  360.00  &   63292  &  -10.27  &   -9.79  &   1.666  &   1.623  \\
  350.00  &   59665  &  -10.40  &   -9.95  &   1.656  &   1.617  \\
 {\bf VII-VIII}\\ 
   69.70  &   55000  &  -11.97  &  -12.01  &   1.719  &   1.718  \\
   69.00  &   70000  &  -11.67  &  -11.71  &   1.758  &   1.758  \\
   69.49  &   80000  &  -11.45  &  -11.48  &   1.781  &   1.782  \\
 {\bf VI-VII} \\
  301.00  &   50742  &  -12.19  &  -10.85  &   1.523  &   1.642  \\
  280.00  &   51549  &  -12.29  &  -11.03  &   1.529  &   1.655  \\
  240.00  &   52454  &  -12.50  &  -11.23  &   1.538  &   1.669  \\
  100.00  &   54767  &  -13.15  &  -11.85  &   1.567  &   1.712  \\
   50.00  &   55369  &  -13.38  &  -12.06  &   1.576  &   1.726  \\

 {\bf VI-VIII} \\
   69.00  &   55000  &  -12.14  &  -12.02  &   1.594  &   1.719  \\
   35.00  &   32974  &  -12.90  &  -12.58  &   1.519  &   1.660  \\

\end{tabular}
\end{table}

\clearpage

\begin{table}
\caption{Triple points of the TIP4P/2005 model of water , between stable phases at high pressures } 
      \label{triple_points}
      \begin{tabular}{lcccc}       \hline
Phase boundary & & T/K  & &  p/bar   \\
\hline
fluid-VI-VII      &  & 301  & & 50742  \\
fluid-VII-plastic &  & 352  & & 60375  \\
VI-VII-VIII       &  & 69   & & 55000    \\
\end{tabular}
\end{table}

\clearpage

\begin{table}
\caption{ Densities of ice VII as obtained from the NpT simulations of this work for the 
TIP4P/2005 and TIP5P models of water.  
The experimental
data for ice VII are taken from Ref. \cite{JCP_1993_99_5369}
and for the rest of polymorphs from Ref. \cite{bookPhysIce}, except for ice II 
that was taken from Ref.\cite{fortes_jac05}. Values of the densities of the other polymorphs 
were taken from Ref.\cite{noya_jpc_2007} for the TIP4P/2005 model and from Ref.\cite{pccpgdr} for the TIP5P model.}
      \label{density_ices}

\begin{tabular}{cccccc}
\hline
Phase & \hspace{0.2cm}T/K & \hspace{0.2cm}p/bar & 
          \multicolumn{3}{c}{$\rho$/(gcm$^{-3})$} \\
    &      &      &Expt   & TIP4P/2005 & \hspace{0.2cm}TIP5P \\
\hline

VII & 300 & 45000 & 1.718 & 1.625 & 1.729 \\
VII & 300 & 70000 & 1.810 & 1.707 & 1.814 \\ \hline
I$_h$ & 250&   0  & 0.920  & 0.921  & 0.976  \\
II  &  123 &   0  & 1.190  & 1.199 & 1.284   \\
III &  250 & 2800 & 1.165  & 1.160 & 1.185   \\
IV  &  110 &   0  & 1.272  & 1.293 & 1.371   \\
V   &  223 & 5300 & 1.283  & 1.272 & 1.331   \\
VI  &  225 & 11000& 1.373  & 1.380 & 1.447   \\
IX  & 165  & 2800 & 1.194  & 1.190 & 1.231  \\
XII & 260  &   5  & 1.292  & 1.296 & 1.340  \\
\end{tabular}
\end{table}

\clearpage

CAPTION TO THE FIGURES

\begin{figure}[!h]
\begin{center}
\caption{a) $f(\theta)$ as a function of $\theta$ angle for ices VII 
(solid line) and VIII (dashed-dotted line) at 300K and 70000bar and for 
the plastic crystal phase (dotted line) at 400K and 70000bar.\\
b) $f(\phi)$ as a function of $\phi$ angle for ices VII (solid line) 
and VIII (dashed-dotted line) at 300K and 70000bar and for the plastic 
crystal phase (dotted line) at 400K and 70000bar. For clarity the function $f(\phi)$ of 
ice VIII has been shifted by 90 degrees to illustrate the similarity with that of ice VII.}
\label{figure_ftheta_vii_viii} 
\includegraphics[clip,height=0.35\textheight,width=0.60\textwidth,angle=-0]{ftheta_vii_viii_fig9a.eps} 
\includegraphics[clip,height=0.35\textheight,width=0.60\textwidth,angle=-0]{fphi_vii_viii_fig9b.eps}
\end{center}
\end{figure}

\end{document}